\documentclass[useAMS,usenatbib]{mn2e}
\usepackage{graphicx} 
\usepackage{times} 
\def\kms{km ${\rm s}^{-1}$}

\def\ch2{$\chi^2$}

\def\Mo{M$_\odot$}
\def\kms {\hbox{${\rm km\ s}^{-1}$}}

\def\scm  {$\hbox{{\rm cm}}^{-2}$}    %cm-2
\def\arcsec {\hbox{$^{\prime\prime}$}}

\def\MOLH {\hbox{${\rm H}_2$}}  %H2
\def \AL {$\alpha $}     %  gr. alpha
\def \HI {H{\sc \,i}}

\def\lapp{\ifmmode\stackrel{<}{_{\sim}}\else$\stackrel{<}{_{\sim}}$\fi}
\def\gapp{\ifmmode\stackrel{>}{_{\sim}}\else$\stackrel{>}{_{\sim}}$\fi}
\def\bsp_small{\vspace{0.5cm}\small\noindent This paper has been typeset
from a \TeX/\LaTeX\ file prepared by the author.\normalsize}

\title[\HI\ and OH absorption towards J0414+0534]{\HI\ and OH absorption in the lensing galaxy of MG J0414+0534}
\author[S. J. Curran et al.]{S. J. Curran$^{1}$\thanks{E-mail: sjc@phys.unsw.edu.au}, 
J. Darling$^{2}$, A. D. Bolatto$^{3,4}$,  M. T. Whiting$^{5}$, C. Bignell$^{6}$, J. K. Webb$^{1}$\\ 
$^{1}$School of Physics, University of New South Wales, Sydney NSW 2052, Australia\\
$^{2}$Center for Astrophysics and Space Astronomy,
Department of Astrophysical and Planetary Sciences,
University of Colorado, 389 UCB,\\ Boulder, CO 80309-0389, USA\\
$^{3}$Department of Astronomy and Radio Astronomy Laboratory, University of
California, Berkeley, CA 94720, USA\\
$^{4}$Department of Astronomy and Laboratory for Millimeter-Wave Astronomy,
University of Maryland, College Park, MD 20742, USA\\
$^{5}$CSIRO Australia Telescope National Facility, PO Box 76, Epping NSW 1710, Australia\\
$^{6}$National Radio Astronomy Observatory, P.O. Box 2, Rt. 28/92
Green Bank, WV 24944-0002, USA}
\begin{document}

\date{Accepted ---. Received ---; in original form ---}

\pagerange{\pageref{firstpage}--\pageref{lastpage}} \pubyear{2007}

\maketitle

\label{firstpage}

\begin{abstract}
We report the detection of \HI\ 21-cm absorption in the $z=0.96$
early-type lensing galaxy towards MG J0414+0534 with the Green Bank
Telescope. The absorption, with total $N_{\rm HI}=1.6 \times
10^{18}\,(T_{\rm s}/f)$ \scm, is resolved into two strong
components, probably due to the two strongest lens components, which
are separated by 0.4\arcsec. Unlike the other three lenses which have
been detected in \HI, J0414+0534 does not exhibit strong OH
absorption, giving a OH/\HI\ column density ratio of $N_{\rm
OH}/N_{\rm HI}\lapp10^{-6}$ (for $T_{\rm s}=100$ K, $T_{\rm x}=10$ K
and $f_{\rm HI}=f_{\rm OH}=1$). This underabundance of molecular gas
may indicate that the extreme optical--near-IR colour ($V-K=10.26$)
along the line-of-sight is not due to the lens. We therefore suggest
that despite the strong upper limits on molecular absorption at the
quasar redshift, as traced by millimetre lines, the extinction occurs
primarily in the quasar host galaxy.
\end{abstract}


\section{Introduction}\label{intro}

Redshifted molecular and atomic absorption systems can provide
excellent probes of the contents and nature of the early Universe. In
particular, with redshifted radio and microwave lines we can
investigate the gaseous content and large-scale structure, as well any
possible variations in the values of the fundamental constants, at
large look-back times. However, such systems are quite rare, with 65
\HI\ 21-cm absorbers at $z\gapp0.1$ currently known, comprising of 37
associated and 28 intervening systems. Three of these occur in
gravitational lenses \citep{cry93,cdn99,kb03}, all of which (PKS
0132--097, B2 0218+35, PKS 1830--211)\footnote{Note that there is also
a possible second lensing system at $z_{\rm abs}=0.19$ towards the
$z_{\rm host}=2.507$ PKS 1830--211, detected in 21-cm \citep{lrj+96} but
not molecular \citep{wc98} absorption.} have also been detected in
molecular absorption, either through the 18-cm OH line
\citep{cdn99,kc02a,kcl+05} or a plethora of different molecules which
absorb in the millimetre regime \citep{wc95,wc98}.

At $z_{\rm host}=2.639\pm0.002$, MG J0414+0534 is a heavily reddened
galaxy \citep{lejt95}, with a known intervening galaxy at $z_{\rm
abs}=0.9584\pm 0.0002$ \citep{tk99}, which is responsible for the
gravitational lensing \citep{sm93}. However, the source of the
reddening, whether in the lens or the host galaxy, is the subject of
much debate: \citet{tk99} argue in favour of the host, as the lensing
galaxy appears as a normal elliptical with a low dust content, while
multi-component spectral analysis by \citet{lejt95} places the dust in
the lensing galaxy (an arrangement similar to PKS 1830--211). This may
be supported by the fact that, while \HI\ absorption with a column
density of $N_{\rm HI}=7.5\pm1.3\times10^{18}.\,(T_{\rm s}/f)$ \scm\ was found
in the host galaxy (at $z_{\rm abs}=2.6365$), HCN was undetected at
$N_{\rm HCN}>10^{13}$ cm$^{-2}$ \citep{mcm98}. This strong limit on
the molecular gas abundance in the host suggests a correspondingly low
dust content, thereby supporting the dusty lens hypothesis.

From extensive millimetre-wave observations of optically selected
objects (see \citealt{cmpw03} and references therein), combined with
the data from the known molecular absorbers, \citet{cwm+06} concluded
that there is a correlation between the OH 18-cm absorption line
strength and the optical--near-IR ($V-K$) colour. This suggests that
the reddening of these background sources is indeed due to dust, as
traced by the molecular abundance. For the systems searched, molecular
fractions of unity are reached for $V-K\gapp5.3$ (figure 1 of
\citealt{cwm+06}). 
MG J0414+0534 has $V-K=10.26$ \citep{lejt95}, and since molecular
absorption was not found in the host galaxy, the lensing galaxy is the
ideal place to search for molecular absorption. In this Letter we
report the results of our search for \HI\ and OH absorption at
$z=0.96$ towards J0414+0534.

% and, since molecular absorption was not found in the
%host, with $V-K=10.26$ \citet{lejt95}, the gravitational lens towards
%J0414+0534 is the ideal target in which to search for molecular absorption
%at high redshift. In this Letter we report the results of our search
%for \HI\ and OH absorption at $z=0.96$ towards J0414+0534.

\section{Observations and Data Reduction}\label{obs}

We observed the redshifted \HI\ 21-cm line at 725~MHz toward
J0414+0534 with the Green Bank Telescope\footnote{The National Radio
Astronomy Observatory is a facility of the National Science Foundation
operated under cooperative agreement by Associated Universities, Inc.}
(GBT) on 22 October, 2006 and 15 February, 2007. The four 18~cm OH
lines were observed simultaneously with the 21-cm line during the 2006
observations. These were conducted in a single 200~MHz bandpass
centered on 810~MHz in 5~min position-switched scans with spectral
records recorded every 5~sec and a winking calibration diode firing
during every other record. The total on-source integration time was
3,280~sec. The autocorrelation spectrometer used 3-level sampling in
two (subsequently averaged) linear polarizations.  Bandpasses were
divided into 32,768 channels and hanning smoothed to 16,384
independent channels for a rest-frame velocity resolution of
5.0~km~s$^{-1}$ in \HI\ and 4.2--4.5~km~s$^{-1}$ in the OH lines.

The 2007 observations were conducted in both 12.5~MHz and 50~MHz
bandpasses in 5~min position-switched scans with spectral records
recorded every 1.5~sec and a winking calibration diode firing during
every other record.  The total on-source integration times were
2,866~sec and 2,686~sec in the two bandpasses, respectively. The
autocorrelation spectrometer used 9-level sampling in two
(subsequently averaged) linear polarizations.  Bandpasses were divided
into 32,768 channels and hanning smoothed to 16,384 independent
channels.  The 12.5~MHz bandpass observations were 4-channel Gaussian
smoothed such that both final spectra had a rest-frame velocity
resolution of 1.26~km~s$^{-1}$.

For all observations, records were individually calibrated and
bandpasses flattened using the calibration diode and the corresponding
off-source records.  Scans and polarizations were subsequently
averaged, and a 5th-order polynomial baseline was fit over a 
3 MHz range (three times that shown in Fig. \ref{HI}) and subtracted.
Systematic flux calibration errors in these data are of order $10\%$.
All data reduction was performed in {\sc
gbtidl}.

The bandpasses were heavily contaminated by radio frequency
interference (RFI) and were far from flat, with the widest bandpass
observations exhibiting the poorest baselines. At 725 MHz, the DC
levels of individual scans showed significant fluctuations and the
levels removed with baseline subtraction in the three final spectra
spanned $3.8\pm1.0$ Jy. At 850 MHz, the continuum level of 4.24 Jy was
not deemed to be reliable and so at this frequency we adopt the flux
density of 3.28~Jy, obtained from Westerbork Synthesis Radio Telescope
(WSRT) observations of this source (see Fig.~\ref{OH}).  Applying a
spectral index of $-1.1$ and using the WSRT value, gives a flux density
of 3.91 Jy at 725 MHz, close to our median value, which we adopt.

\section{Results and Discussion}
\subsection{Observational results}

\begin{figure}
\vspace{7.6cm}
\includegraphics{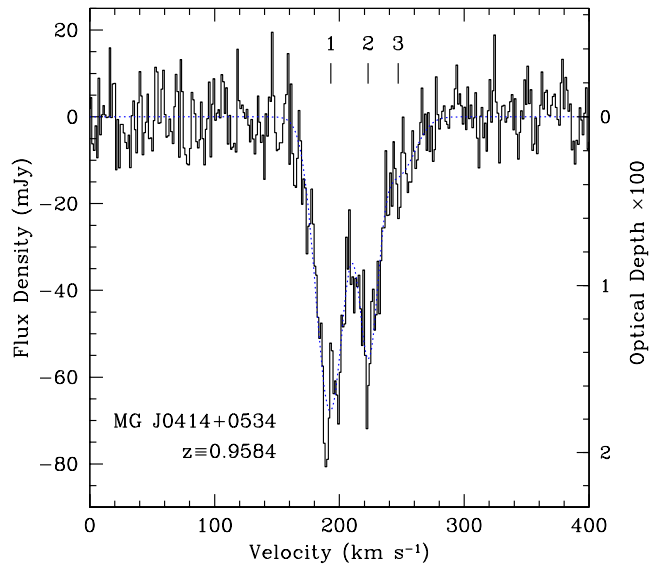}
\caption{\HI\ spectrum of the MG J0414+0534 lens from the 2007
  12.5~MHz bandpass session (which had the highest spectral
  resolution, flattest bandpass and lowest r.m.s. noise). The velocity
  scale is in the lens rest frame, using the optical redshift
  of $z=0.9584$. The left hand ordinate shows the observed flux
  density after baseline subtraction and the right hand ordinate shows
  the observed optical depth for a continuum of 3.91 Jy. The dotted line shows the 3-component Gaussian fit with the
  labels referring to the centroid of each component.  These and the
  other line properties are given in Table \ref{t1}.}
\label{HI}
\end{figure}

In Figs. \ref{HI} and \ref{OH} we show the reduced spectra of the \HI\
and OH observations, respectively, where we see that for the \HI\ detection a
two component fit is required, with possibly a third required for the weak
redshifted wing.
\begin{table*}
 \centering
 \begin{minipage}{150mm}
  \caption{\HI\ absorption properties at $z=0.96$ towards MG
J0414+0534 derived from the Gaussian fits. $\nu$ [MHz] is the observed
frequency (barycentric, optical definition), $z_{\rm abs}$ the
corresponding redshift, followed by the resulting \HI-optical rest
frame velocity offset [\kms], FWHM is
the full-width half maximum of the line [\kms], followed by the observed 
peak depth [mJy].
Combined with the flux density of $S=3.91$ Jy, this gives the observed
optical depth, which in the optically thin regime is related to the actual 
optical depth via $\tau=f\,\tau_{\rm act}$,  where $f$ the
covering factor of the background continuum source.  
The last two
columns give the velocity integrated depth [\kms] and the derived column
density [\scm], where $T_{\rm s}$ is the spin temperature [K]. The quoted
uncertainties are from the Gaussian fits and the bottom row describes
the aggregate line properties obtained without the fitting of individual
components. \label{t1}}
  \begin{tabular}{@{}cccccccc c@{}}
  \hline
Line & $\nu$  & $z_{\rm abs}$ & $v_{\rm HI} - v_{\rm opt}$ & FWHM  & Depth  &$\tau$ &$\int\tau\,dv$ & $N_{\rm HI}$ \\
  %&  (MHz)  &             & (km s$^{-1}$) & (kHz) & (km s$^{-1}$) & (mJy)&  &(km s$^{-1}$) \\
%(1)&(2)&(3)&(4)&(5)&(6)&(7)&(8)&(9)\\
 \hline
1 & 724.823(1) & 0.959658(4) & +193(31)  & 27(1) & $-$68(2) & 0.0176(5) & 0.50(2)   & $9.1\times10^{17}.\,(T_{\rm s}/f)$\\
2 & 724.749(2) & 0.959858(6) & +223(31)  & 18(2) & $-$51(7) & 0.0131(18)& 0.25(4)   & $4.6\times10^{17}.\,(T_{\rm s}/f)$\\
3 & 724.692(18)& 0.960013(49)& +247(31)  & 31(14)& $-$14(2) &  0.0036(5) & 0.12(6)  & $2.2\times10^{17}.\,(T_{\rm s}/f)$\\
\hline
Total&724.792(2)&0.959743(6) & +205(31)  & 49(2) & $-$82(6) & 0.0212(16)& 0.88(2)   & $1.60(4)\times10^{18}.\,(T_{\rm s}/f)$\\
\hline
\end{tabular}
\end{minipage}
\end{table*}
\begin{figure}
\vspace{7.7cm}
%\vspace{7.5cm}
%\special{psfile=MG0414Icc.eps hoffset=15 voffset=0 hscale=43 vscale=43 angle=0}
\includegraphics{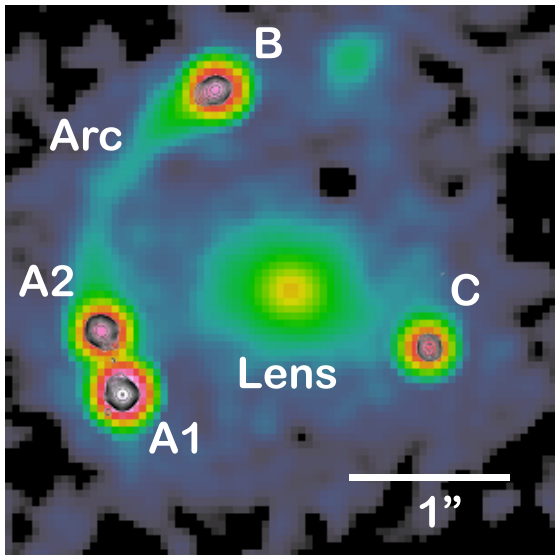}
\caption{$I$-band image of the lens and lensed components (A1, A2, B
\& C) of J0414+0534 \citep{fls97} [+CASTLES;``cleaned'' image], in
colour, overlaid with the ($\geq3\sigma$) 22 GHz contours from
\citep{kmh97}, in black. This frequency exhibits the radio emission at
a similar resolution to the optical image, where the radio contours 
are exactly coincident with the peak optical emission of each component.}
\label{lens}
\end{figure}
In Fig. \ref{lens}, we show the $I$-band image of J0414+0534, overlaid with
high resolution radio contours. A1, A2, B \& C are the lensed images
of the background $z_{\rm host}=2.64$ quasar, with the arc being the
more extended galaxy hosting this. The arc exhibits no radio emission and
is less obscured than the nuclear region hosting the AGN (see
Sect. 3.2). The two main \HI\ absorption peaks are probably due to A1 \&
A2, with the wide shallow absorption (3 in Table/Fig. \ref{t1}),
possibly associated with these main components or with the weaker B or
C components. Due to their projected
distance\footnote{$H_{0}=75$~km~s$^{-1}$~Mpc$^{-1}$, $\Omega_{\rm
matter}=0.27$ and $\Omega_{\Lambda}=0.73$.} of up to $\approx15$ kpc
from the main components, while maintaining a velocity difference of
$\lapp60$ \kms\ (Table \ref{t1}), it is unlikely that B or C are
responsible for the redshifted wing apparent in the profile.

%These calculations assume the gas is contained in a rotating disk associated with the lens galaxy. 
%However, it is unlikely that a
It is surprising that a 
passive early-type galaxy such as the lens galaxy would exhibit such
high column densities (Table~\ref{t1}) in its outskirts. The
significant velocity
offset between the \HI\ and the optical galaxy is suggestive of gas
infalling to the lens galaxy (although the gas could be on the far
side of the lens and be flowing away from the galaxy). There are
several galaxies seen in the surrounding field \citep{avhm94} [albeit
with no measured redshift], so there is also the possibility of a
group interaction generating the gas streams.

If we assume, however, that the \HI\ is bound to the lensing galaxy in
an inclined disk, and that the absorption arises just from components
A1 \& A2, we can set restrictions on the location of the absorbing gas.
%Since, for all three components, there is a significant difference in
%velocity from the lensing galaxy's optical redshift, we can
%Alternatively, we can 
%set restrictions on the location of the absorbing gas by
%assuming that the gas lies in an inclined \HI\ disk, and the
%absorption arises just from components A1 \& A2. 
The radio emission
from these components lies at a projected distance of 10.1~kpc from the
lens galaxy's centre, just beyond the Einstein radius \citep{twh00}.
\citet{fls97} showed that this lens follows a deVaucouleurs surface
brightness profile with an ellipticity of 0.20(2) and a position angle
of 71$^\circ$(5). We find that there is no disk inclination that can
reproduce the observed velocity gradient unless the position angle of
the lens differs substantially ($\sim20^\circ$) from that of the putative
disk. 

Assuming gravitational motion, we estimate the mass enclosed by the
\HI\ components 1 and 2 (as source images A1 and/or A2) from
%\begin{equation}
\[
 M(r<10.1~{\rm kpc}) = {R\,(v_{\rm HI}-v_{\rm opt})^2\over G\, \sin^2\theta\,\cos\phi},
%\end{equation*}
\]
where $\theta$ and $\phi$ are the projection angles of the gas
velocity vector and source image-lens distance,
respectively\footnote{$\theta=\phi=0$ for velocity and distance
vectors lying in the plane of the sky.}. Hence, modulo unknown
projection corrections, from the optical redshift of the lens $M_1
\geq 8.7(1.4)\times 10^{10}$~M$_\odot$ and $M_2 \geq 11.6(1.6)\times
10^{10}$~M$_\odot$, cf. the $10.7\times10^{10}$~M$_\odot$ obtained
from the Einstein ring radius of \citet{rgm+00}.  The dominant
statistical uncertainty is the optical redshift of the lens (rest
frame $\approx30$ \kms, Sect. 1) and the dominant systematic
uncertainties are the unknown projections of the \HI\ velocity vectors
and the lens-gas separation.  These could easily boost the enclosed
mass values by an order of magnitude if the plane of motion is close
to face-on.  Note that the total lens mass obtained from models by
\citet{fsw05} within 29.0~kpc (well beyond the Einstein radius) is
$82.5^{+11.4}_{-7.2} \times 10^{10}$~M$_\odot$.  If the gas resides in
a dark matter halo with a flat rotation curve, then our mass estimates
would be a factor of $\sim3$ larger at this radius.  It is likely that
$\phi$ is small and $\cos\phi\sim1$; in this case, we require only
that $\theta\lapp35^\circ$ to match our mass estimates to those of
\citet{fsw05}.

\begin{figure}
\vspace{7.7cm}
\includegraphics{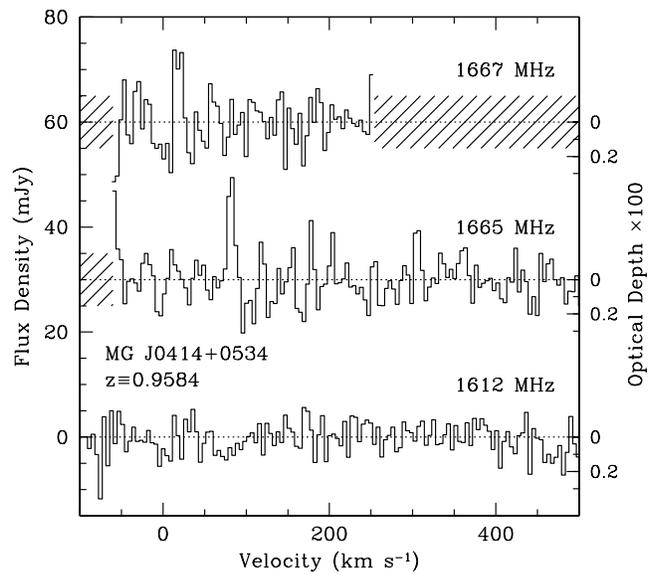}
\caption{OH 18~cm non-detection spectra of the MG J0414+0534 lens. As
per Fig. \ref{HI}, the velocity scale is in the lens rest frame and
the right hand ordinate shows the observed optical depth for a
continuum of 3.28~Jy. The 1720~MHz line
was unobservable due to RFI, and the cross-hatching shows ranges in
the other bands which were also lost. The derived limits are given in
Table \ref{t2}.  Note that the 1665 and 1667 MHz lines were also
undetected in WSRT observations, which reached an r.m.s. of 14 mJy per
each 7 \kms\ channel over the redshift range $z=0.9410-0.9826$ [1667
MHz] (Curran et al. in prep).}
\label{OH}
\end{figure}
 \begin{table*}
 \centering
 \begin{minipage}{130mm}
  \caption{OH absorption properties at $z=0.96$ towards MG
J0414+0534. The parameters are as Table \ref{t1}, but $\delta v$ [\kms] is
the channel spacing and $\sigma$ [mJy] is the r.m.s. noise per channel.  The
last three columns are derived for the $3\sigma$ limit per $\delta v$
channel using a flux density of 3.28 Jy, where $T_{\rm x}$ is the
excitation temperature [K].\label {t2}}
  \begin{tabular}{@{}ccccccc c@{}}
  \hline
Line & $\nu$  & $z_{\rm abs}$ & $\delta v$  & $\sigma$  &$\tau$ &$\int\tau\,dv$  & $N_{\rm OH}$ \\
 \hline
1612 MHz & 822.0--823.4 & 0.9580--0.9614 & 4.4 & 2.6 & $<0.0021$& $<0.0093$& $<2.0\times10^{13}.\,(T_{\rm x}/f)$\\
1665 MHz & 848.6--850.1 & 0.9591--0.9625 & 4.3 & 4.1 &$<0.0038$ & $<0.016$& $<6.9\times10^{12}.\,(T_{\rm x}/f)$\\
1667 MHz & 850.7--851.3 & 0.9586--0.9600 & 4.3 & 5.2 &$<0.0048$ & $<0.020$& $<4.9\times10^{12}.\,(T_{\rm x}/f)$\\
\hline
\end{tabular}
%{\bf  after smoothing the 1612 line may give the best limit on N(OH). RESULT IS ``SMOOTHED'' IN NEXT SECTION, BUT IT WOULD BE NICE TO SEE A SPECTRUM OF THIS}
\end{minipage}
\end{table*}

\subsection{The source of the reddening}

\begin{figure}
\vspace{7.6 cm} \setlength{\unitlength}{1in}
\begin{picture}(0,0)
\put(-0.2,3.6){\includegraphics{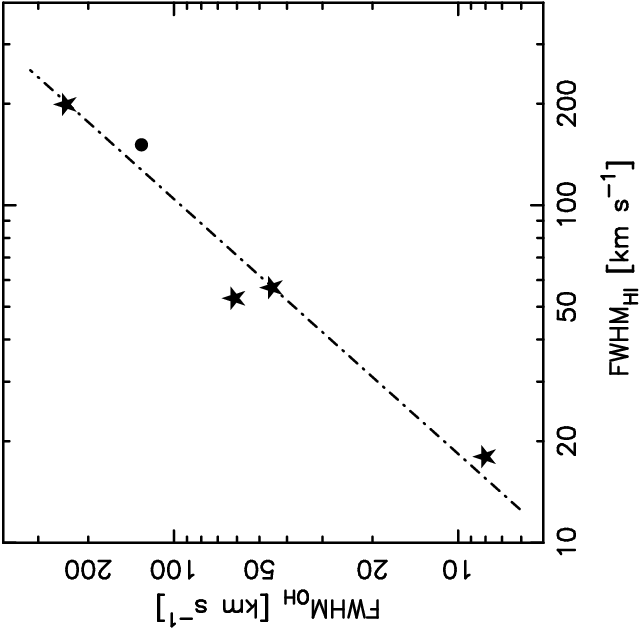}}
\end{picture}
\caption{The approximate 1667 MHz OH full width half maxima versus those of the \HI\ for
the known OH absorption systems (from
\citealt{cps92,cmr+98,cdn99,kc02a,kb03,kcdn03,kcl+05}). The stars
represent the four millimetre absorption systems and the circle
represents 0132--097 where OH, but no millimetre, absorption has been
detected. The least-squares fit has a regression coefficient of 98.1\%.
This correlation may suggest that the atomic and molecular gases
are spatially coincident, cf. the atomic and singly ionised
gases in damped Lyman-\AL\ absorbers \citep{ctp+07}.}
\label{vel}
\end{figure}

Our 1667 MHz OH limit (Table \ref{t2}), gives a $3\sigma$ velocity
integrated optical depth limit of $\int\!\tau_{\rm OH}\, dv < 0.020$
\kms, per each 4.3 \kms\ channel. However, the known 1667 MHz line
profiles are generally much wider than this, therefore in order to obtain a
meaningful column density limit, the optical depth limit must
be integrated over a similar range. Since the line has not
been detected, we have no knowledge of its width, although we do find
a correlation between the \HI\ and OH line widths for the known
systems (Fig. \ref{vel}). So from our \HI\ detection, we estimate the
OH line width to be FWHM$_{\rm OH}\approx40$ \kms.
Using this value significantly reduces (by a factor of 9) the
sensitivity in terms of column density, although some is recovered by
the fact that our data are at a much higher resolution than that
required to detect such a wide line. Therefore for a comparable
estimate of our limit, we multiply the velocity integrated optical
depth by $\sqrt{{\rm FWHM}_{\rm OH}/\delta v}$, giving the limit for a
single channel ``smoothed'' to FWHM$_{\rm OH}$. This gives
$\int\!\tau_{\rm OH}\, dv < 0.06$ \kms\ ($3\sigma$), resulting in a
normalised OH line strength of $N_{\rm OH}/N_{\rm HI}\lapp9
\times10^{-6}\,.\,(f_{\rm HI}/f_{\rm OH})\,.\, (T_{\rm x}/T_{\rm s})$.
\begin{figure}
\vspace{7.6 cm} \setlength{\unitlength}{1in}
\begin{picture}(0,0)
\put(-0.2,3.6){\includegraphics{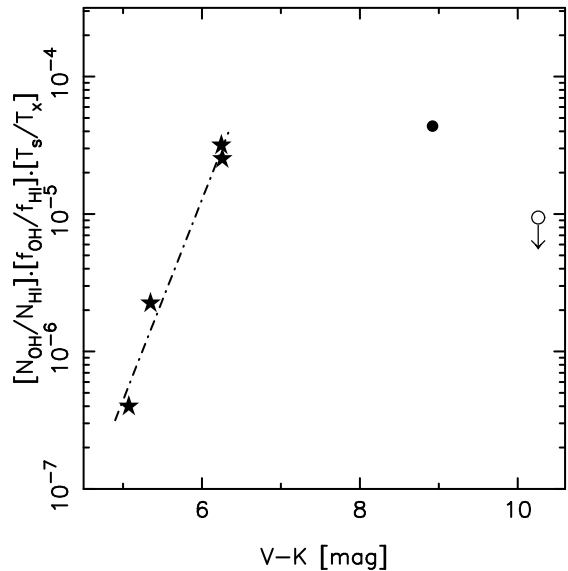}}
\end{picture}
\caption{The normalised OH line strength ($2.38\times10^{14}\int
\tau_{_{\rm OH}}\, dv/1.82\times10^{18}\int\tau_{_{\rm HI}}\, dv$)
versus optical--near-IR colour. The symbols are as per Fig.~\ref{vel},
with the unfilled circle designating the upper limit for J0414+0534.
The least-squares fit to the millimetre absorption systems has a
regression coefficient of 98.6\%.
Note that correlation is not just limited to this range, with
\MOLH-bearing damped Lyman-\AL\ absorption systems exhibiting
molecular fractions of $\sim10^{-7}-10^{-2}$ and colours of $V-K=2.2 -
3.4$ (figure 1 of \citealt{cwm+06}).}
\label{OHoverH}
\end{figure}

\citet{cwm+06} found a correlation between the normalised OH line
strength and the $V-K$ colour (see Fig.~\ref{OHoverH}). If this is
reliable,
%If the correlation between the normalised OH line strength and $V-K$ is
%reliable, this 
the very low strength for J0414+0534 % (Fig.~\ref{OHoverH})
suggests that the dust causing the large amount of reddening is not
coincident with the \HI\ we see in absorption, and so is not
located in the lens. The low HCN abundance, $N_{\rm
HCN}\leq1.2\times10^{13}$ cm$^{-2}$ (for $T_{\rm x}=10$
K),
%\footnote{Using $T_{\rm x}=17$ K (i.e. $T_{\rm x}=10$ K at $z=0$) gives $N_{\rm HCN}\leq3.2\times10^{13}$ cm$^{-2}$ (http://www.phys.unsw.edu.au/$\sim$sjc/column/).}, 
at $z=2.6365$
\citep{mcm98} appears to argue against the presence of significant
quantities of dust in the host galaxy. However, this may not be the
case.  We note that a non-detection of a millimetre line does not
necessarily imply an extremely low molecular abundance: Until the
detection of OH absorption in the $z_{\rm abs}=0.765$ gravitational
lens towards the $z=2.225$ ``quintuple quasar'' PKS 0132--097 (PMN
J0134--0931), the HCO$^+$--to--OH column density ratio of $N_{\rm
OH}\approx 30N_{\rm HCO^+}$ for Galactic clouds \citep{ll96} was found
to hold for all four of the known redshifted OH absorbers
\citep{kc02a}. However, although \citet{kcl+05} detected OH at a
column density of $N_{\rm OH}=5.6\times10^{14}.\,(T_{\rm x}/f)$ \scm,
HCO$^+$ $J=1\rightarrow2$ was undetected to $\tau<0.07$ at a $3\sigma$
level. For a channel spacing of 5~\kms, this gives $N_{\rm
HCO+}\leq1\times10^{12}$ cm$^{-2}$ (for $f=1$ and $T_{\rm x}=10$ K),
i.e. $N_{\rm OH}\gapp 5600 N_{\rm HCO^+}$.

For the four known millimetre absorbers, $N_{\rm HCN}\approx 0.4 - 3
N_{\rm HCO^+}$ \citep{wc95,wc96,wc96b,wc97} and so a similar limit to
that in 0132--097 could allow OH column densities of $N_{\rm
OH}\lapp6\times10^{15}.\,(T_{\rm x}/f)$ \scm, or a normalised OH line
strength of $\lapp0.01$ (cf. Fig. \ref{OHoverH}), in the nucleus of
J0414+0534. The fact that a decimetre (OH) absorption could be
detected, while millimetre (HCN) is not (as in 0132--097), could be
due to geometry: A molecular cloud has a much larger chance of
occulting the lower frequencies, since the emission region is larger,
in fact close to the same size as the 21-cm emission region. However,
for the more compact millimetre emission, the same molecular cloud has
a much smaller likelihood of intervening this, and so, although OH may
be readily detected, millimetre transitions are not necessarily seen
along our line-of-sight.

From the lack of OH absorption in the lens we therefore suggest that
most of the extinction does not occur here, but is intrinsic to
the nuclear regions of the quasar host. This supports the results of:
\begin{itemize}
  \item \citet{fls97}, who find that the optical arc, due to 
	the more extended background galaxy (Fig. \ref{lens}),
	is significantly bluer ($R-I=1.3$) than the A1/A2 \&  B images
	of the central AGN. Furthermore, \citet{avc+99} find A2 to be the 
	most obscured component. Although this could be due
        to the lens, the bluer arc again suggests that the differential
	extinction occurs across the AGN and its host galaxy.

    \item \citet{tk99}, who find that if the dust were in the lens,
      its distribution would have to be very uniform over the
      $\approx14$ kpc span of the lensed images.  Furthermore, a high
      extinction by the lens galaxy would double the galaxy's luminosity,
      to the point were it would be an anomaly.

   \item \citet{eht+06}, who, from an analysis of the extinction curves,
      find the extinction to be high for
     an early-type galaxy and add the further possibility that some of the extinction is due to an, as yet, unidentified foreground object.
\end{itemize}
Note also that CO emission has also been detected at the host
redshift, indicating a molecular gas mass of up to $M_{\rm
H_2}\sim2\times10^{11}\,m^{-1}\,h^{-2}_{70}$ \Mo, where $m$ is the
lensing magnification factor \citep{baga98}. Finally, the 21-cm
velocity integrated optical depth in the host is $\approx 5$ times
higher than in the lensing galaxy (giving $N_{\rm HI}=1.6$,
cf. $7.5\times10^{18}.\,(T_{\rm s}/f)$ \scm, Table \ref{t1} and
\citealt{mcm98}, respectively). If the neutral gas in the lens and host share
similar spin temperatures and covering factors, this would imply that
$N_{\rm HI_{\rm host}}\sim5\, N_{\rm HI_{\rm lens}}$, indicating
denser gas, although in the absence of any knowledge of these two
parameters such a conclusion is uncertain.

\section{Summary}

We have detected \HI\ 21-cm, but not OH 18-cm, absorption in the
$z=0.96$ gravitational lens towards MG J0414+0534. The \HI\ profile is
resolved into three components, at least two of which we believe are
due to the lens components A1 \& A2. From this, we estimate that the
mass of the lens enclosed by the components is $\gapp10^{11}$ \Mo,
consistent with previous results at low inclinations.

Although there was no detection, upon demonstrating a correlation
between the full-width half maxima of the \HI\ and OH profiles of
other high redshift absorbers,we estimate that any OH associated with
this system would have a FWHM$_{\rm OH}\approx40$~\kms. The $3\sigma$
optical depth limit of $\tau<0.020$ per 4.3 \kms\ for the 1667MHz OH
line, therefore translates into a minimum detectable OH column density
of $N_{\rm OH}\approx1.5\times10^{13}.\,(T_{\rm x}/f)$, giving a
normalised OH line strength of $N_{\rm OH}/N_{\rm HI}\lapp9
\times10^{-6}\times(f_{\rm HI}/f_{\rm OH})(T_{\rm x}/T_{\rm s})$.
      
According to the correlation of the molecular fraction with the
optical--near-IR colour for the known redshifted molecular absorption
systems \citep{cwm+06}, this extremely low $N_{\rm OH}/N_{\rm HI}$
ratio supports much of the published literature, which does not favour
the lens as the cause of the extreme red colour of the background
$z_{\rm host}=2.64$ quasar. Although no molecular absorption at
millimetre wavelengths was detected in the host, this does not
preclude this as the source of the reddening, since,
as in the case of PKS 0132--097, geometrical effects may be
at play. This could be tested by searching for OH absorption
in the host of J0414+0534.

\section*{Acknowledgments}

We would like to thank the reviewer for their prompt and helpful
report and Nissim Kanekar for the velocity integrated OH optical depths
towards PKS 0132--097. This research has made use of the NASA/IPAC
Extragalactic Database (NED) which is operated by the Jet Propulsion
Laboratory, California Institute of Technology, under contract with
the National Aeronautics and Space Administration.  This research has
also made use of NASA's Astrophysics Data System Bibliographic
Services.

%Use no bold or italic, no commas after author surnames, and no ampersand between the final two author names. List all authors if eight or fewer, otherwise first author only followed by `et al.'.

%\bibliographystyle{mn2e} % FOR VERY FINAL VERSION

%\bibliographystyle{apj} %same problem as before (m.tex)
%\bibliography{aa,ref}
%\expandafter\ifx\csname natexlab\endcsname\relax\def\natexlab#1{#1}\fi

\label{lastpage}
\end{document}